\documentclass{aastex62}
\usepackage{float}
\usepackage{dsfont}
\usepackage{gensymb}

\usepackage{amsmath}

\definecolor{orange}{RGB}{255,127,0}

\graphicspath{{./}{Figs/}}

\submitjournal{ApJL}

\shorttitle{Heat flux regulation in the expanding solar wind}
\shortauthors{Innocenti et al.}

\begin{document}

\title{Collisionless heat flux regulation via electron firehose instability in presence of a core and suprathermal population in the expanding solar wind}
\correspondingauthor{Maria Elena Innocenti}
\email{mariaelena.innocenti@kuleuven.be}

\author{Maria Elena Innocenti}
\affil{University of Leuven (KULeuven), Department of Mathematics, Centre for mathematical Plasma Astrophysics, Celestijnenlaan 200B, Leuven, Belgium}

\author{Elisabetta Boella}
\affiliation{Lancaster University, Physics Department, Bailrigg, Lancaster LA1 4YW, UK}
\affiliation{Cockcroft Institute, Sci-Tech Daresbury, Keckwick Lane, Warrington WA4 4AD, ,UK}

\author{Anna Tenerani}
\affiliation{Department of Physics, The University of Texas at Austin, TX 78712}

\author{Marco Velli}
\affiliation{University of California Los Angeles, 
Department of Earth, Planetary, and Space Sciences, 
595 Charles E Young Dr E, 
Los Angeles, CA 90095}
\begin{abstract}

The evolution of the electron heat flux in the solar wind is regulated by the interplay between several effects: solar wind expansion, that  can potentially drive velocity-space instabilties, 
turbulence and wave-particle interactions, and, possibly, collisions. Here we address the respective role played by the solar wind expansion and the electron firehose instability, developing in the presence of multiple electron populations, in regulating the heat flux. We carry out fully kinetic, Expanding Box Model simulations and separately analyze the enthalpy, bulk and velocity distribution function skewness contributions for each of the electron species.  We observe that the key factor determining electron energy flux evolution is the reduction of the drift velocity of the electron populations in the rest frame of the solar wind. In our simulations, redistribution of the electron thermal energy from the parallel to the perpendicular direction after the onset of the electron firehose instability is observed. However, this process seems to impact energy flux evolution only minimally. Hence, reduction of the electron species drift velocity in the solar wind frame appears to directly correlate with efficiency for heat flux instabilities.

\end{abstract}

\keywords{instabilities, plasmas, methods: numerical, Sun: heliosphere, solar wind}

\section{Introduction} \label{sec:intro}

Electrons play a fundamental role in solar wind dynamics, 
because they drive coronal expansion and carry the greatest share of the heat flux, due to their light mass~\citep{feldman75}. A comprehensive solar wind theory then requires understanding the mechanisms behind electron heat flux regulation in the expanding solar wind. Outstanding open questions are whether collisional, Spitzer-H{\"a}rm~\citep{spitzer1953transport, salem2003electron} or collisionless~\citep{scime1994regulation, crooker2003suprathermal, landi2012competition} processes (the transition between regimes is discussed, e.g., in~\citet{bale2013electron} and references therein) act as main heat flux regulator and, in the latter case, which specific collisionless process (i.e., instability) plays the most prominent role. Candidates instabilities 
are proposed (or ruled out) in~\citet{gary1975electron, scime1994regulation, gary1994whistler, scime2001solar, roberg2018wave, komarov_schekochihin_churazov_spitkovsky_2018, tong2019whistler, vasko2019whistler, verscharen2019self, lopez2019particle, kuzichev2019nonlinear}.

The closer one gets to the Sun, the more the electron energy flux problem intersects the coronal heating problem, and the fundamental issue of solar wind acceleration~\citep{scudder1983collapse, meyer1999does, dorelli1999electron, lie200116, landi2001temperature, dorelli2003electron}. 

The role of collisions in heat flux regulation falls within the larger topic of their role in solar wind evolution
~\citep{lie1997kinetic, lie200116, landi2012competition, Yoon2019_PRL}. In approaching the discussion, one should keep in mind that collisions and wave-particle interactions ultimately act in the same direction, i.e. reducing the anisotropy of velocity distributions driven by the solar wind expansion, and that higher than observed collisionality levels may in fact approximate the effects of wave-particle interactions in models that, ``on paper", should not account for them. 

Solar wind expansion plays a fundamental role in heat flux evolution. Even in the simplified (and unrealistic) double adiabatic (DA) framework~\citep{chew1956boltzmann}, where wave-particle interactions and dissipative processes are neglected, one can expect a drop of the electron energy flux with heliocentric distance
~\citep{scime1994regulation}. Abandoning the DA framework in favor of a more realistic solar wind description, where wave/particle interaction occurs, 
 the possibility arises that expansion modifies solar wind bulk parameters in a way that facilitates the onset of wave/particle resonances and kinetic instabilities that can, in turn, regulate the heat flux. 

Solar wind expansion increases the parallel beta and reduces the perpendicular to parallel temperature ratio: this drives the system towards ion~\citep{hellinger2003hybrid, matteini2006parallel} and electron~\citep{innocenti2019onset} firehose instabilities, which in fact are observed to constrain the ion and electron populations in the respective ``Brazil'' plots~\citep{matteini2007evolution, matteini2013signatures, vstverak2008electron, bervcivc2019scattering}.

The Electron Firehose Instability (EFI) is an electromagnetic kinetic instability that develops in the presence of a background magnetic field and of a $T_{\bot}< T_{\parallel}$ 
thermal anisotropy 
over spatial and temporal scales that are relatively large and slow for electrons, and exhibits lower threshold and higher growth rates at oblique, rather than parallel, propagation~\citep{li2000electron, paesold2000electron,  gary2003resonant, camporeale2008electron}. 

Studies of the EFI are usually (with few exceptions, e.g.~\citet{shaaban2018clarifying, shaaban2019firehose}) limited to a single, non drifting, electron population, which therefore does not carry significant heat flux in the mean velocity frame. 
To understand heat flux dynamics, instead, one has to take into account the \textcolor{black}{multiple electron populations which} compose the electron Velocity Distribution Function (eVDF) in the solar wind.

In this paper, we discuss global electron heat flux regulation by the EFI triggered as a result of solar wind expansion within a purely collisionless description of the solar wind. We simulate plasma expansion self-consistently with the semi-implicit, fully kinetic Expanding Box Model (EBM) code EB-iPic3D~\citep{innocenti2019semi, innocenti2019onset}. The EBM framework maps the evolution of a solar wind plasma parcel that moves radially away from the Sun with constant velocity $U_0$, while expanding in the transverse directions with characteristic expansion timescale $\tau= R/U_0$ ($R$ is the heliocentric distance), to a Cartesian, non-expanding, co-moving grid where the secular evolution with distance appears through time-dependent terms and coefficients. In EB-iPic3D,  such grid is resolved with a fully kinetic, semi-implicit algorithm~\citep{markidis2010multi, innocenti2017progress}. Other EBM implementations rely on hybrid~\citep{liewer2001alfven, matteini2006parallel, hellinger2013protons}
or MagnetoHydroDynamics~\citep{tenerani17_AEB} descriptions.

\textcolor{black}{The solar wind eVDF is composed of three populations: a colder, nearly Maxwellian core, a tenuous, suprathermal  halo, and a field aligned strahl ~\citep{feldman75, pilipp87, maksimovic2005radial, Horaites2018}. Since the aim of this work is to study heat flux evolution, this eVDF can be simplified with a two population eVDF composed of a core component and a tenuous, suprathermal population (representing both the halo and the strahl) which we call ``suprathermal electrons". The heat flux resulting from such an eVDF presents characteristics similar enough to the observed heat flux (see Section~\ref{sec:HF} later) to justify the approximation.} 

In Section~\ref{sec:EFI}, we investigate how plasma expansion drives our two-electron-component (``2E") simulation into the EFI-unstable area. For comparison, we evolve also a one-electron-component  (``1E") simulation. In Section~\ref{sec:HF} we then focus on the role of the EFI in regulating the heat-flux in the presence of multiple electron populations. The formula of~\citet{feldman75} is used to distinguish among enthalpy, bulk, and heat flux in the frame of reference of the species components within the total electron energy flux.

\section{Electron Firehose Instability onset}
\label{sec:EFI}
We compare the development of the oblique EFI in the 2E and 1E simulations. 
Our simulation setup is quite similar to~\citet{lopez2019particle}. We simulate a 1D box with length $L_x/d_i= 16$, with $d_i$ the ion skin depth, resolved with 1024 cells. 
The mass ratio is $m_r=1836$ and the Alfv\'{e}n speed is $v_A/c= 0.00023$, with $c$ the speed of light: both electron skin depth and the electron gyroradius are resolved. The time step is $\omega_{pi}t= 0.0375$, with $\omega_{pi}$ the ion plasma frequency. The ratio between the plasma frequency and the gyrofrequency is $\omega_{pi}/ \Omega_{ci}= 4390$ and $\omega_{pe}/ \Omega_{ce}= 102$ respectively for ions and electrons, compatibly with values observed in the solar wind. 
The magnetic field at initialization is inclined with an angle $\theta_0=26^{\circ} $ with respect to the x, radial direction. In the 2E simulation, 
the core (``c") and \textcolor{black}{suprathermal (``s")} populations are modeled with Maxwellian distributions with thermal velocity $v_{th,c}/c= 0.0119512$, $v_{th,s}/c= 0.0292744$ and density $n_c/n_e=0.95$, $n_s/n_e=0.05$ respectively, with $n_e$ the total electron density. Their drift velocity in the radial, x direction is $V_{x,c}/c=-0.000479551 $ and $V_{x,s}/c=0.00911147$, which satisfies the zero-current relation~\citep{feldman75, scime1994regulation}. The ions have the same temperature as the core electrons, $v_{th,i}/c= 0.00027$. In the 1E simulation, 
the entire electron population has $v_{th,e}/c= 0.0119512$ and is not drifting with respect to the ions. 5000 particles per species per cell are used in all cases.

The EB expansion time, $\tau$, is $ \omega_{pi} \tau= R_0/ U_0 \sim 5500$, with $R_0/d_i= 20$ the initial distance of the box from the Sun, and $U_0/c= 0.00358536$ the solar wind velocity. In EB simulations $R_0$ should be regarded as a sort of free parameter, to choose in order to have expansion dynamics fast enough to impact the evolution of the simulation, but slow with respect to the characteristic timescales of the processes under investigation (in this case, EFI evolution), as assumed in the formulation of the EB method. In our choice of simulation parameters we respect the solar wind ordering $v_A \sim v_{th,i} << U_0 < v_{th,e}$.

\begin{figure}[ht]
        \centering
        
            \includegraphics[width=0.475\textwidth]{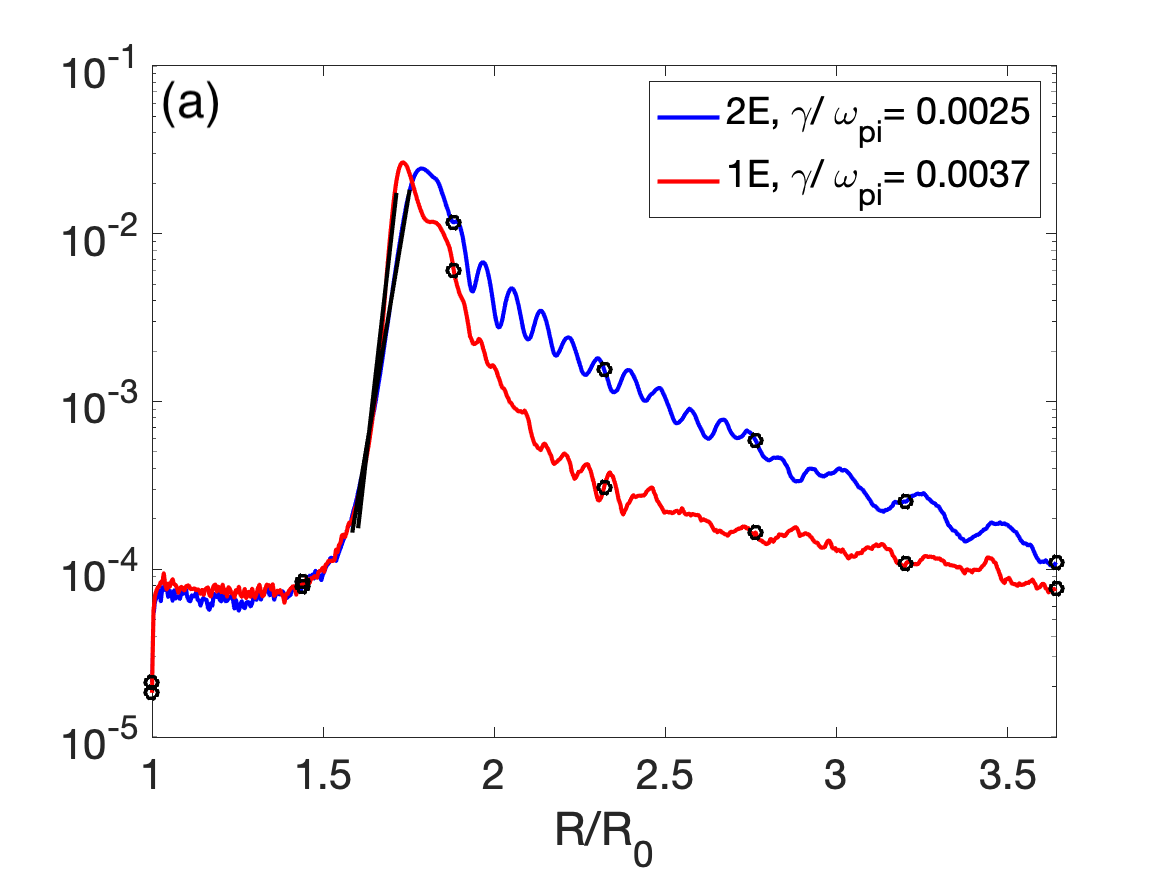}
            \includegraphics[width=0.475\textwidth]{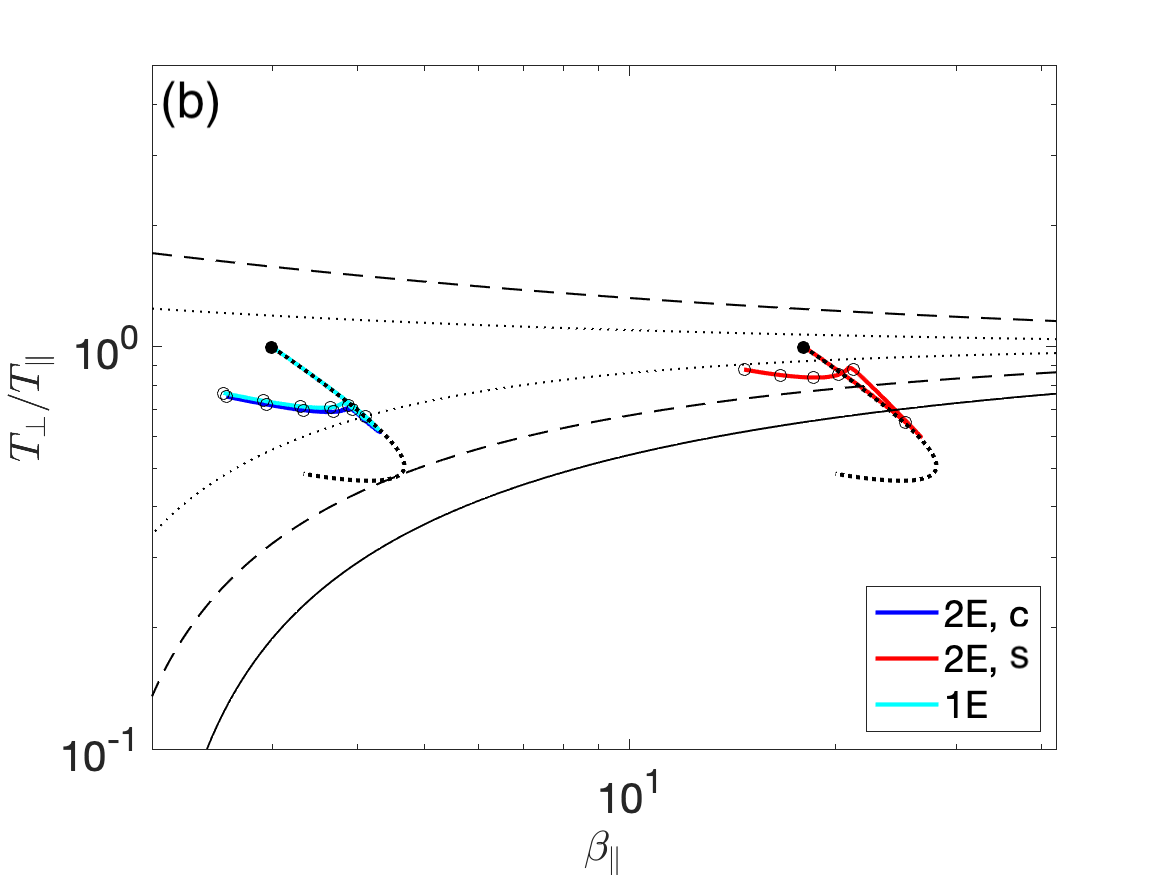}
            \includegraphics[width=0.475\textwidth]{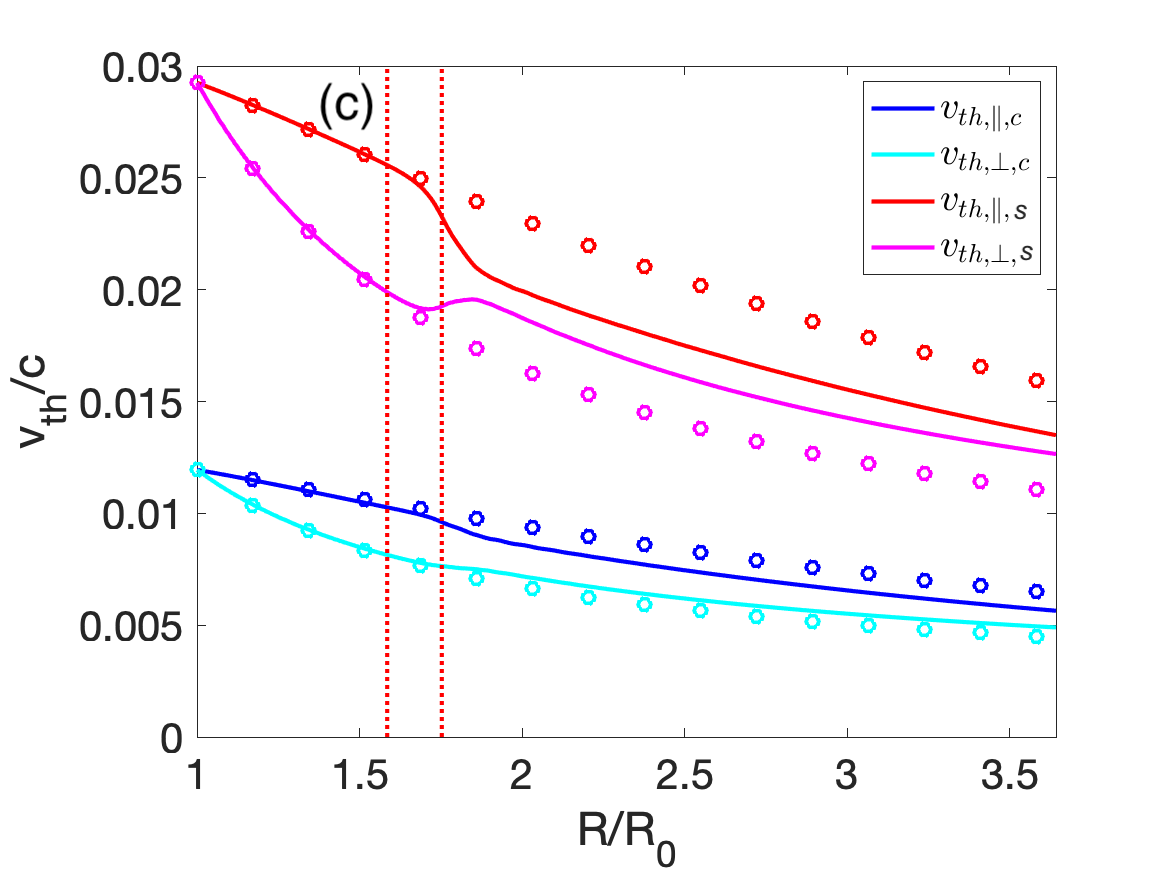}
            \includegraphics[width=0.475\textwidth]{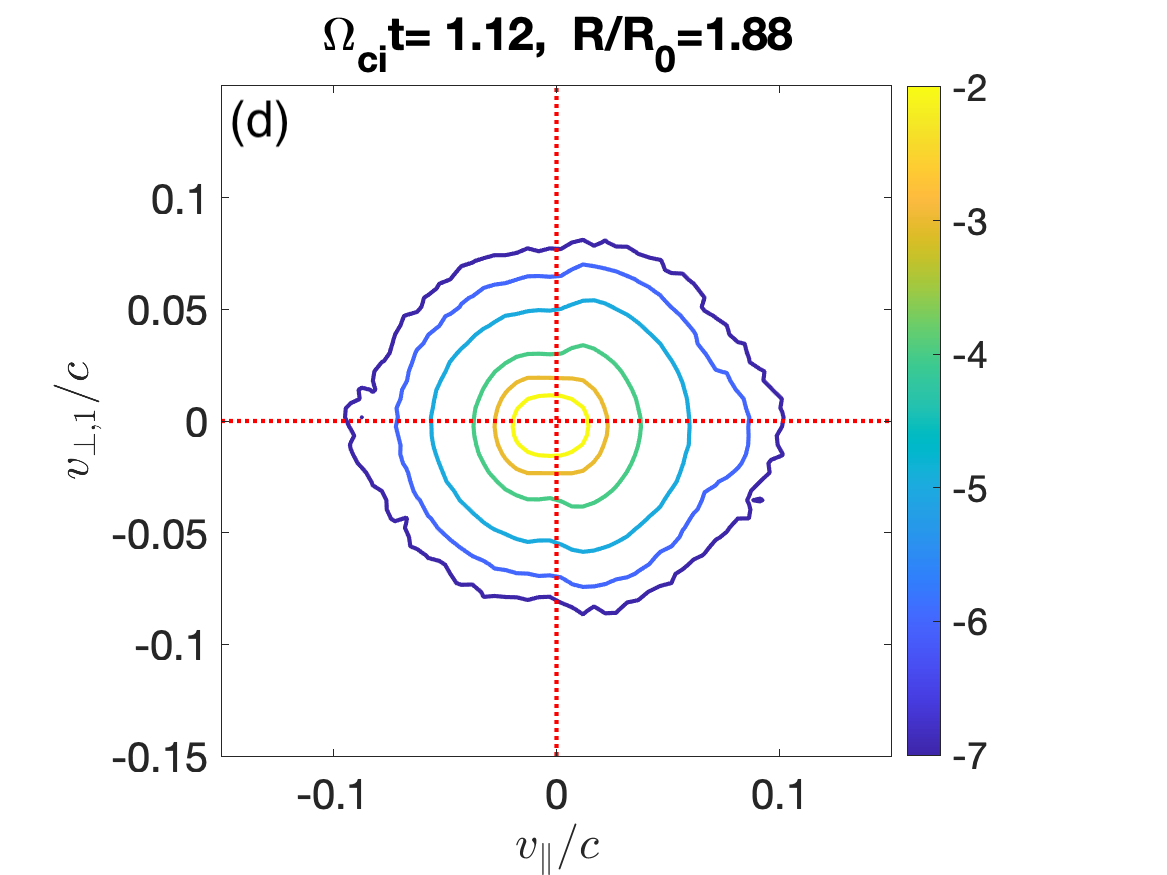}
           
        \caption
        {\small (a): Oscillating magnetic field energy normalized to the mean magnetic field energy for the 2E and 1E simulations, 
        as a function of $R/R_0$.  (b): traces of the core and \textcolor{black}{suprathermal electrons} (blue and red) in the 2E simulation, of the entire electron population in the 1E simulation in the electron $\beta_{\parallel}$ vs $T_{\bot}/ T_{\parallel}$ plot. The initial time is marked with a black, filled dot. 
        The resonant firehose and whistler isocontours are from~\citet{gary2003resonant} and~\citet{GaryWang_1996_whistler}. 
        Empty black dots are drawn in correspondence of the heliocentric distances where eVDFs are plotted in the movies. (c): evolution of the parallel and perpendicular thermal energy of the core (blue and cyan) and \textcolor{black}{suprathermal} (red and magenta) electrons in solid lines, compared with the expansion-only driven evolution, in circles, for the 2E simulation. The vertical lines mark the linear growth rate phase of the EFI.  
        (d) eVDF in the parallel vs perpendicular direction for the 2E, simulation at $R/R_0= 1.88$, $\Omega_{ci}t= 1.12$ (third dot in panel a and b), immediately after the end of the EFI linear growth phase. In the videos added as supplemental material, eVDF evolution during the entire simulations for the 2E and 1E cases. 
        }  
\label{fig:global}
\end{figure}

In Figure~\ref{fig:global}, panel a, we show the oscillating magnetic field energy for the 2E (blue) and 1E (red) simulations. The global evolution of the magnetic and kinetic energy, in the two simulations, matches remarkably well the double adiabatic expectations. Observing the evolution of the oscillating magnetic field energy (which constitutes a small fraction of the total magnetic energy), we notice, in both simulations, the onset of an oblique EFI with growth rate $\gamma/ \omega_{pi}= 0.0025$ and $\gamma/ \omega_{pi}= 0.0037$ respectively, for the case with two and one electron species. We notice that these growth rates give a ratio between the expansion time scale $\tau$ and the EFI e-folding time $t_{e-f}$ of $\tau/ t_{e-f} \sim 14 $ and $\tau/ t_{e-f} \sim 20 $, well in line with the $\tau > t_{e-f}$ requirement of EB models.

The identification of the developing instability as an oblique EFI is supported by Figure~\ref{fig:global}, panel b, where we plot the traces of the two simulations in the electron $\beta_{\parallel}$ vs $T_{\bot}/ T_{\parallel}$ plane. The blue and red traces are the core and \textcolor{black}{suprathermal}  electrons in the 2E simulation, the cyan trace corresponds to the electrons in the 1E simulation. 
The traces ``bounce back" in the stable area of the plot after briefly entering the EFI-unstable area. This is a departure from the ``ideal expansion-driven traces", depicted with dotted lines, which show how the simulation would evolve due to expansion alone, i.e. in the absence of instability development. 
The differences between the 2E and 1E simulation traces are minimal, since the density \textcolor{black}{of the suprathermal electrons} is very low when compared to the core density in the 2E simulation. 
Figure~\ref{fig:global}, panel c, shows the thermal velocity evolution of the core and \textcolor{black}{suprathermal electrons}, parallel and perpendicular components, as a function of $R/R_0$ compared with expansion-driven expected behavior. We see that the observed evolution (solid lines) matches well the expansion driven one (circles) before the onset of the instability. After that, energy is redistributed from the parallel to the perpendicular direction. 

We depict in Figure~\ref{fig:global}, panel d, the eVDF at $R/R_0= 1.88$ (third dot in panel a and b), immediately after the end of the linear growth phase of the EFI. In the videos uploaded as supplemental material, the eVFD evolution during the entire simulation for the 2E and 1E simulations is provided. 

In the video, we see a first phase of double-adiabatic-like cooling. Since the initial magnetic field is not purely radial, the cooling affects both the parallel and the perpendicular direction. At $R/R_0= 1.88$, consistently with the oscillating magnetic field energy evolution shown in Figure~\ref{fig:global}, panel a, the 1E eVDF starts showing traces of resonant wave-particle interaction, similar to those observed in~\citet{innocenti2019onset} in 2D3V EBM simulations. The core \textcolor{black}{electrons} of the 2E simulation exhibits similar traces. When the entire electron population of the 2E simulation is plotted, two features are quite evident: the reduction of the initial drift between the core and the \textcolor{black}{suprathermal} component, and the emergence of an asymmetry between the $v_{\parallel}<0$ and $v_{\parallel}>0$ sides of the eVDF.

The former feature is associated with a reduction of the observed heat flux, as discussed in the next section.

\section{Heat Flux Regulation}
\label{sec:HF}

In the 2E simulation, the asymmetry of the initial eVDF in the $v_{x}<0$ and $v_{x}>0$ semi-planes originates an electron heat flux, which evolves with the eVDF.

In our simulation, the rest frame of the entire electron population, core plus \textcolor{black}{suprathermal electrons}, corresponds, due to the zero-current relation, to the co-moving frame, i.e. the rest frame of the solar wind.
The total electron heat flux (that is also the total electron energy flux in this frame ~\citep{feldman75}) is composed of a core and \textcolor{black}{suprathermal} contribution.
Labeling the electron populations, core and \textcolor{black}{suprathermal}, with $j$, the heat flux associated with population $j$, $\mathbf{Q}_j$, 
is composed of three terms~\citep{feldman75}:
\begin{equation}
    \mathbf{Q}_{j} = m_e /2 \int \mathbf{v}v^2 f_j d^3v= 
    n_j \mathbf{V}_{d,j} (3/2 T_{\parallel,j} + T_{\bot,j}) + m_e/ 2 \; n_j \mathbf{V}_{d,j} V_{d,j}^2+ \mathbf{q}_j = \mathbf{Q}_{enth,j} + \mathbf{Q}_{bulk,j}+ \mathbf{q}_j,
    \label{eq:Feld_energyFlux}
\end{equation}
where $\mathbf{v}$ is the velocity and $\mathbf{V}_{d,j}$ the species drift velocity in the co-moving frame. The heat flux \textit{in the frame of reference of each population},  $\mathbf{q}_{j}$, is 
\begin{equation}
    \mathbf{q}_{j} = m_e /2 \int \mathbf{w}_j w_j^2 f_j d^3v
    \label{eq:Feld_heatFlux}
\end{equation}
with $\mathbf{w}_j= \mathbf{v}- \mathbf{V}_{d,j}$. $\mathbf{Q}_{enth,j}$ and $\mathbf{Q}_{bulk,j}$ are the enthalpy and bulk velocity contributions. While $\mathbf{Q}_{enth,j}$ and $\mathbf{Q}_{bulk,j}$ depend on ``large scale" solar wind parameters, such as parallel and perpendicular temperatures and drift velocities, $\mathbf{q}_j$ is a direct measure of the skewness of the VDF. In~\citet{scime1994regulation}, it is noted that Eq.~\ref{eq:Feld_energyFlux} excellently reproduces Ulysses energy flux observations between 1 and 5 AU.

\begin{figure}[ht]
        \centering
          \includegraphics[width=0.475\textwidth]{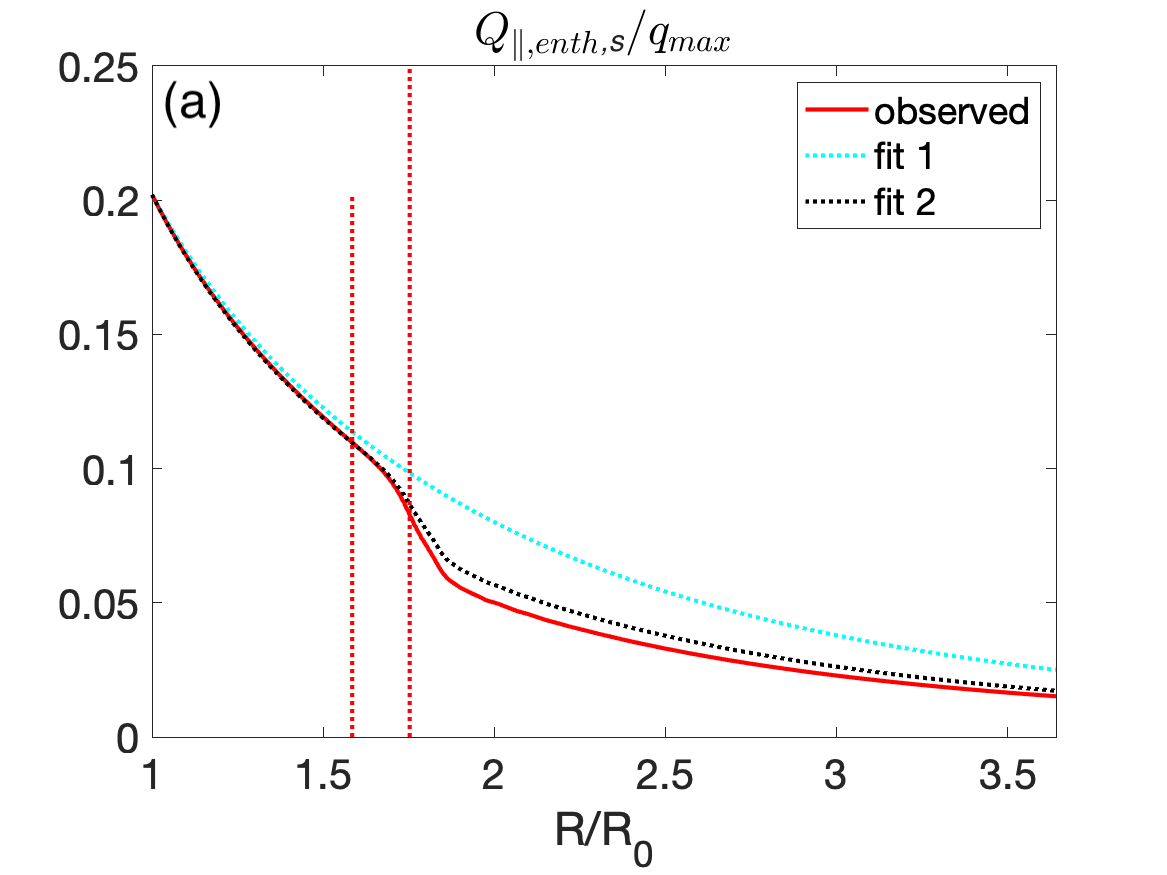}
         \includegraphics[width=0.475\textwidth]{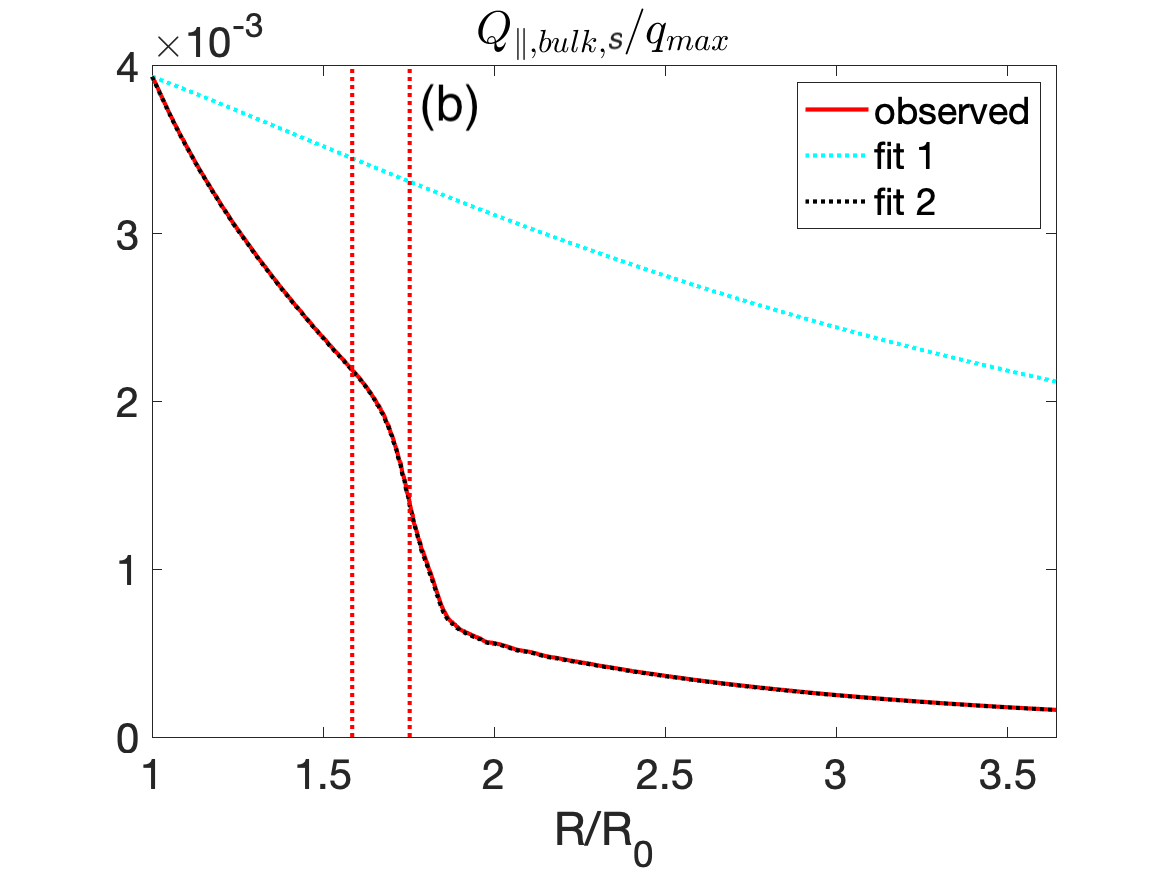}
         \includegraphics[width=0.475\textwidth]{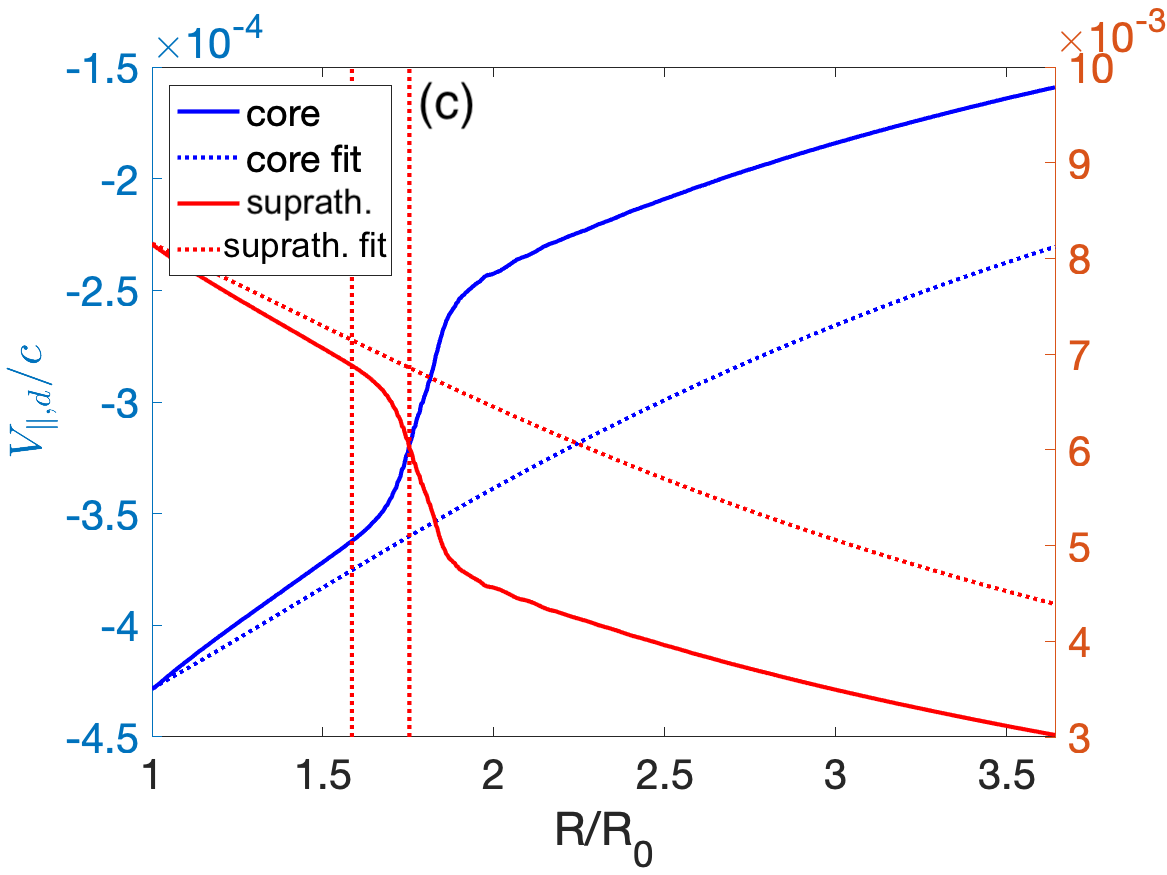}
         \includegraphics[width=0.475\textwidth]{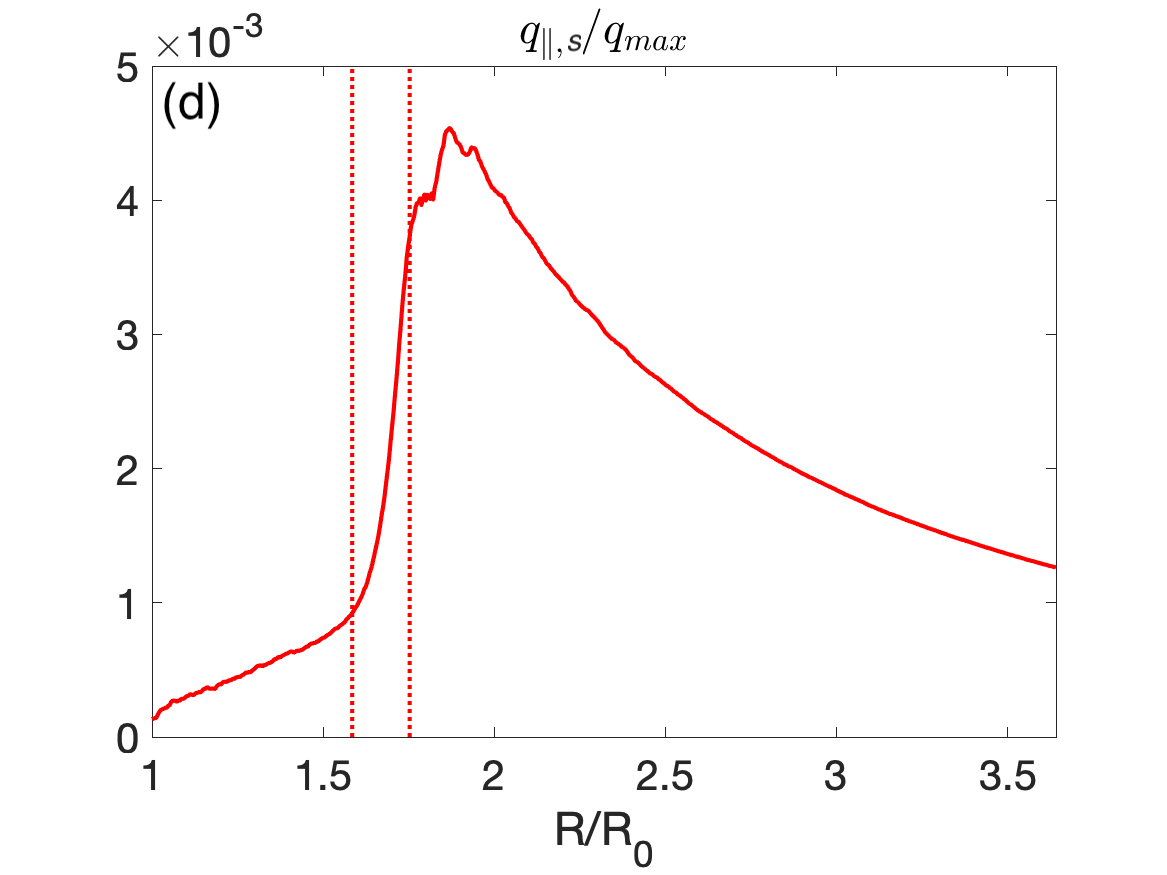}
        \caption
        {\small panel (a) and (b): Evolution of the parallel \textcolor{black}{suprathermal electron} enthalpy (panel a) and bulk (panel b) components, as observed from the simulation. The cyan and black lines are fits calculated using the expansion driven evolution for the thermal velocities and the expansion driven values (cyan line) or observed values (black line) for the parallel drift velocity. panel (c): Evolution of the core (blue, left axis) and \textcolor{black}{suprathermal electrons} (red, right axis) parallel drift velocity. The dotted lines are the expansion driven fits. panel (d): Evolution of the parallel \textcolor{black}{suprathermal} heat flux. The vertical lines mark the linear growth rate phase of the EFI. } 
\label{fig:HF_fits}
\end{figure}

We observe that the \textcolor{black}{suprathermal electrons} carry the largest contribution to the energy flux, and that the two population contributions are oppositely directed, with the \textcolor{black}{suprathermal} contribution being anti-Sunwards~\citep{feldman75, scime1994regulation}. Since the velocity drifts are initialized in the radial, rather than parallel, direction, smaller perpendicular energy flux \textcolor{black}{components} (not depicted here, and generally negligible in the solar wind) are observed together with the dominant parallel \textcolor{black}{ones}.
In Figure~\ref{fig:HF_fits}, we plot in red the \textcolor{black}{suprathermal} enthalpy, bulk and heat flux $q_s$ (panel a, b, and d) contributions to the parallel textcolor{black}{suprathermal} energy flux $Q_s$,  customarily normalized to the free streaming value $q_{max}=3/2 m_e \sum_j n_j v_{th,j,0}^3$ (where we have considered that the two electron species have different thermal velocities).

In the simulations as in the solar wind, the convection of the \textcolor{black}{suprathermal} electron enthalpy, $Q_{\parallel,enth,s}$, largely dominates the energy flux~\citep{feldman75, scime1994regulation} (compare the y axis scales in panel a, b, d).

The electron energy flux decreases during the simulation due to a combination of three processes: plasma expansion, reduction of the relative drift velocity of the populations with respect to the solar wind rest frame, onset and development of the EFI. 

To disentangle the role of these three processes, we plot in Figure~\ref{fig:HF_fits}, panel a and b, two fits for the parallel enthalpy and bulk components. The first fit, in cyan, is calculated from Eq.~\ref{eq:Feld_energyFlux} using as thermal velocities 
the DA evolution 
(depicted as circles in Figure~\ref{fig:global}, panel c). In this fit, the module of the drift velocity is kept constant, and the parallel component is calculated projecting the initial drift velocity value, in the x direction, in the direction of the background magnetic field at each heliocentric distance, as calculated from DA 
evolution (since the radial magnetic field component drops with $R$ faster than the transverse component, the angle of the magnetic field vector with the radial direction increases with $R$). 
For the second fits, in black, we use expansion driven thermal velocity evolution (as before) and, as drift velocity, the values calculated at each heliocentric distance from the simulation.

We observe in both cases that the ``black" fit excellently reproduces the observed values. The ``cyan" fit is poorer. In the particular case of the enthalpy plot, panel a, we notice that the cyan fit is adequate only up to EFI onset.
This means that solar wind expansion alone (cyan fit), while contributing to the observed enthalpy and bulk component evolution, does not completely explain it. Collisionless processes, reproduced within our fully kinetic, expanding box simulation, reduce the heat flux to values lower than what expected from expansion effects alone, as observed in the solar wind already in~\citet{scime1994regulation}.

In Figure~\ref{fig:HF_fits}, panel c, we depict 
the parallel component of the drift velocity for the core (blue, left axis) and the \textcolor{black}{suprathermal electrons} (red, right axis), as a function of $R/R_0$ and marking with vertical lines the linear growth phase of the EFI. In dotted line we depict the 
parallel drift component as expected from DA 
evolution 
(the dotted lines were used to calculate the cyan fits). 
We notice that the sudden drops in the drift velocities 
are associated to the 
EFI. The EFI thus affects enthalpy and bulk components chiefly  through its effects on the species drift velocities, rather than by redistributing energy from the parallel to the perpendicular direction (the expansion-driven thermal velocity evolution, which does not account for EFI energy redistribution, still gives a good fit of the observed values, black fits in Figure~\ref{fig:HF_fits}, panel a and b). 
Finally, in Figure~\ref{fig:HF_fits}, panel d, we plot the parallel \textcolor{black}{suprathermal electron} heat flux in the rest frame of the \textcolor{black}{suprathermal electrons}, $q_{\parallel,s}$.
We notice that its value increases, as a consequence of EFI onset, due to the eVDF asymmetry in the parallel direction highlighted in Figure~\ref{fig:global}, panel d. After saturation of the EFI, the heat flux starts decreasing, seemingly following the trend of the magnetic field oscillations.


\section{Discussion}
\label{sec:concl}

In this paper, we analyze the role of solar wind expansion and selected collisionless processes (namely, the electron firehose instability in presence of two electron populations) in electron heat flux regulation. Our key finding is that, at least in our simplified scenario, the drift velocity between the electron populations plays a fundamental role in solar wind electron energy flux regulation. In our simulation, we observe a thermal velocity evolution that differs from double adiabatic expectations due to EFI onset. However, this does not seem to crucially affect heat flux regulation. The electron energy flux is instead regulated by the sudden drop of the electron relative drift velocity due to EFI onset. Solar wind plasma expansion indirectly contributes to the heat flux regulation, by triggering the EFI instability. 
This work supports an indirect role of solar wind expansion in electron heat flux regulation, where expansion drives or modifies the evolution of heat flux instabilities. 

\textcolor{black}{We notice that the suprathermal (halo plus strahl) population could have been more realistically modelled with a non-Maxwellian distribution, composed e.g. of a kappa distribution for the halo and of the realistic shape for the strahl described in~\citet{Horaites2018}. We expect that simulations initialized with such a VDF would not differ significantly from the one presented here, since EFI onset is driven by the core (see Figure~\ref{fig:global}, panel b), and therefore is not affected by the shape of the suprathermal eVDF. Also, drift velocity evolution of the suprathermal electrons is driven by the halo, since it has higher fractional density than the strahl~\citep{maksimovic2005radial}. The halo can be modeled with a kappa distribution, which is expected to behave similarly to a Maxwellian for processes that do not depend on the high energy tails of the distribution.}

\citet{scime1994regulation} study how the electron energy flux, obtained from Ulysses observations, varies between 1 and 5 AU. They obtain a dependence with the heliocentric distance $r$ of $\propto r^{-2.7}$, inconsistent with the expected variation from collisional processes alone and steeper than what is expected from expansion-driven evolution alone: they deduce that collisionless processes should contribute to heat flux regulation.

They confirm the validity of the phenomenological formula of~\citet{feldman75}, rewritten here as Eq.~\ref{eq:Feld_energyFlux}, in explaining energy flux evolution with heliocentric distance, and confirm the~\citet{feldman75} intuition that the collisionless mechanisms (i.e., instabilities) responsible for heat flux regulation are the ones that also control drift velocity evolution. 

An obvious limitation of the present work is the fact that the global temperature profiles before EFI onset follow a double-adiabatic evolution, rather than profiles with heliocentric distance consistent with solar wind observations - as shown, for example, in~\citet{landi2012competition}. This is due to our simplified initial conditions, that do not allow for processes, such as turbulence, that we believe affect the observed radial profile of the temperature at least as much as collisions. \textcolor{black}{Incidentally, we observe that introducing turbulent fluctuations in the initialization fields may result in a more accurate evolution of the eVDF. In~\citet{tang2018numerical}, for example, it is observed that introducing whistler wave turbulence into the kinetic Fokker-Planck transport equation allows to obtain a core-halo-strahl eVDF and realistic radial dependences for key quantities starting from a core and a suprathermal electron component, with the latter being scattered by whistler fluctuations.}

In future work, we will address this issue and analyze other scenarios, prone to the development of other instabilities potentially controlling energy flux evolution through drift velocity regulation, for a better understanding of the outer corona energy balance. The aim is to provide a reliable interpretative framework for observations of heat flux at different heliocentric distances along the same magnetic field line, which will become available with coordinated Parker Solar Probe~\citep{fox2016solar} and Solar Orbiter~\citep{mueller2013solar} observational campaigns.

\acknowledgments

M.E.I. thanks Rodrigo L\'{o}pez for useful discussions. M.E.I.'s work is supported by an FWO postdoctoral fellowship. The simulations were performed on the supercomputer Marconi-Broadwell (Cineca, Italy) under a PRACE allocation. A. T. acknowledges support from NASA HSR grant N. 80NSSC18K1211.


\end{document}